\documentclass[conference]{IEEEtran}
\IEEEoverridecommandlockouts
\usepackage{cite}
\usepackage{amsmath,amssymb,amsfonts}
\usepackage{algorithmic}
\usepackage{graphicx}
\usepackage{textcomp}
\usepackage{xcolor}
\usepackage{subcaption}
\usepackage{listings}
\usepackage{enumitem}
\usepackage[ruled]{algorithm2e}
\usepackage{caption}
\newcommand{\changed}[1]{\textcolor{black}{#1}}
\DeclareCaptionFont{8pt}{\fontsize{8pt}{6pt}\selectfont}
\def\BibTeX{{\rm B\kern-.05em{\sc i\kern-.025em b}\kern-.08em
    T\kern-.1667em\lower.7ex\hbox{E}\kern-.125emX}}
\begin{document}

\makeatletter % changes the catcode of @ to 11
\newcommand{\linebreakand}{%
  \end{@IEEEauthorhalign}
  \hfill\mbox{}\par
  \mbox{}\hfill\begin{@IEEEauthorhalign}
}
\makeatother % changes the catcode of @ back to 12

\title{Performance Comparison of DAOS and Lustre for Object Data Storage Approaches}

\author{
\IEEEauthorblockN{Nicolau Manubens}
\IEEEauthorblockA{\textit{European Centre for Medium-Range} \\
\textit{Weather Forecasts}\\
Bonn, Germany \\
nicolau.manubens@ecmwf.int}
\and
\IEEEauthorblockN{Simon D. Smart}
\IEEEauthorblockA{\textit{European Centre for Medium-Range} \\
\textit{Weather Forecasts}\\
Reading, United Kingdom \\
simon.smart@ecmwf.int}
\and
\IEEEauthorblockN{Tiago Quintino}
\IEEEauthorblockA{\textit{European Centre for Medium-Range} \\
\textit{Weather Forecasts}\\
Reading, United Kingdom \\
tiago.quintino@ecmwf.int}
%\and
%\IEEEauthorblockN{Emanuele Danovaro}
%\IEEEauthorblockA{\textit{European Centre for Medium-Range} \\
%\textit{Weather Forecasts}\\
%Reading, United Kingdom \\
%emanuele.danovaro@ecmwf.int}
\linebreakand
\IEEEauthorblockN{Adrian Jackson}
\IEEEauthorblockA{\textit{EPCC, The University of Edinburgh}\\
Edinburgh, United Kingdom \\
a.jackson@epcc.ed.ac.uk}
}

\maketitle

\begin{abstract}

High-performance object stores are an emerging technology which offers an alternative solution in the field of HPC storage, with potential to address long-standing scalability issues in traditional distributed POSIX file systems due to excessive consistency assurance and metadata prescriptiveness.

In this paper we assess the performance of storing object-like data within a standard file system, where the configuration and access mechanisms have not been optimised for object access behaviour, and compare with and investigate the benefits of using an object storage system.

Whilst this approach is not exploiting the file system in a standard way, this work allows us to investigate whether the underlying storage technology performance is more or less important than the software interface and infrastructure a file system or object store provides.

%Novel object storage solutions potentially address long-standing scalability issues with POSIX file systems, and Storage Class Memory (SCM) offers promising performance characteristics for data-intensive use cases. Intel’s Distributed Asynchronous Object Store (DAOS) is an emerging high-performance object store which can leverage SCM and NVMe devices. It has been gaining traction after scoring top positions in the I/O 500 benchmark.

%Numerical Weather Prediction (NWP) simulations are sensitive to I/O performance and scaling, and their output resolution and diversity is expected to increase significantly in the near future.

%In this work, we present a preliminary assessment of DAOS in conjunction with SCM on a research HPC system and evaluate its potential use as HPC storage at a world-leading weather forecasting centre. We demonstrate DAOS can provide the required performance, with bandwidth scaling linearly with additional SCM nodes in most cases, although choices in configuration and application design can impact achievable bandwidth. We describe a new I/O benchmark and associated metrics that address object storage performance from application-derived workloads that can be utilised to explore real-world performance for this new class of storage systems. 
\end{abstract}


\section{Introduction}
\changed{Object stores are a candidate to address long-standing scalability issues in POSIX file systems, including excessive consistency assurance and prescriptiveness\cite{nextplatform_lockwood}.}

Numerical Weather Prediction (NWP) usually entails object-like data access, \changed{as global weather fields are currently of the order of 1 MiB in size, relatively small if compared to traditional high-performance I/O sizes}, and an advanced indexing mechanism is required for high-performance semantic discovery and access, which involves several metadata operations. Domain-specific object stores have been developed to implement this semantic indexing on traditional distributed file systems in a way that satisfies current operational NWP performance requirements\cite{fdb-pasc19}.

\changed{With the planned resolution increases in NWP simulations, resulting in one to two orders of magnitude larger data sets, and the advent of general-purpose high-performance object stores, which are specially designed for the type of object-like operations common in NWP, adapting the domain-specific store currently in use at ECMWF becomes an increasingly appealing pathway}. In a recent study, as prior research to validate such effort, we have assessed the performance that DAOS, a high-performance object store which has been recently gaining traction, can provide together with Storage Class Memory (SCM) when tested with an ad-hoc benchmark which mimics I/O patterns in our operational NWP use case\cite{DAOS-IEEETPDS-ARXIV}.

In this paper we review these DAOS performance results, and compare with corresponding performance results obtained using Lustre, one of the most popular distributed file systems in HPC, after adapting the ad-hoc benchmark to carry out the object operations on top of a file system.

This work allows us to discriminate the benefits achieved from using specific storage hardware as compared to the benefits from the object store design and implementation. It also allows us to draw conclusions of general interest on the benefits of object stores, and gives some real use-case same-hardware same-software data points for comparison of Lustre and DAOS.

\section{DAOS}

The Distributed Asynchronous Object Store (DAOS)\cite{daos-scfa2020} is an open-source high-performance object store designed for \changed{massively distributed non-volatile memory (NVM)} including SCM and NVMe. It provides a low-level key-value storage interface on top of which other higher-level APIs, also provided by DAOS, are built. Its features include transactional non-blocking I/O, fine-grained I/O operations with zero-copy I/O to SCM, end-to-end data integrity and advanced data protection. The OpenFabrics Interfaces (OFI) library is used for low-latency communications over a wide range of network back-ends.

DAOS is deployable as a set of I/O processes or engines, one per physical socket in a server node, each managing access to SCM and NVMe devices within the socket. An engine partitions the storage it manages into targets to optimize concurrency, each target being managed and exported by a dedicated group of threads. DAOS allows reserving space distributed across targets in so-called \textit{pools}, a form of virtual storage. A pool can serve multiple transactional object stores called \textit{containers}.\changed{A container is a private object address space, which can be modified transactionally and independently of the other containers in the same pool. An application first needs to connect to the pool and then open the desired container. If successfully authorised, the application obtains a handle it can use for its processes to interact with the container.}

Upon creation, objects in a container are assigned a 128-bit unique object identifier, of which 96 bits are user-managed. Objects can be configured for replication and striping across pool targets by specifying their \textit{object class}. \changed{An object configured with striping is stored in parts, distributed across targets, enabling concurrent access.}

\section{Lustre}

Lustre\cite{lustre-arxiv2019} is the foremost parallel file system used at HPC site globally. It is an open-source file system, that aims to provide high bandwidth and high availability for many users across a wide range of hardware. Lustre provides a POSIX-compliant interface to distributed data storage that enables large numbers of clients to connect and use the file system concurrently.

\section{Methodology}

In this work, DAOS and Lustre have been deployed on the NEXTGenIO research HPC system, \changed{exploiting the same underlying hardware, and the deployments have been benchmarked with the community-developed IOR benchmark\cite{ior_repo} and the Field I/O benchmark.}

In brief, the Field I/O benchmark consists of a pair of functions that perform writing and reading of weather fields to and from a DAOS object store, using the DAOS C API. Their design closely mimics the domain-specific object store already employed within ECWMF, and they can be combined and run in parallel in different ways, resulting in two different data access patterns of interest:

\begin{itemize}[leftmargin=*]
    \item \textit{pattern A}: in a first phase, writer programs are run on a number client nodes (typically more than one per node), and issue a sequence of write operations. Once all writers have finished, a second phase runs, where an equal number of reader programs are executed, issuing a sequence of corresponding read operations. This pattern aims to assess the maximum write or read throughput the storage can provide to applications, mimicking a scenario where the NWP writer applications are run separately from the post-processing reader applications.
    \item \textit{pattern B}: firstly, the storage is pre-populated with some data. Following this initialisation step, half of the client nodes employed for the benchmark issue a sequence of write operations, while the other half issue corresponding read operations. This pattern aims to assess the throughput storage can provide in more realistic scenarios with contention between writing and reading processes.
\end{itemize}

Due to its design, Field I/O is representative of the types of I/O workload exhibited from real NWP workflows. Typically, operational NWP workflows at ECMWF operate as in pattern B, with approximately 250 HPC nodes writing simulation output to storage while another 250 nodes read and run post-processing tasks. However many users also run the forecasting models or post-processing applications independently in a manner that is equivalent to pattern A.

Field I/O is described in more detail in\cite{DAOS-IEEETPDS-ARXIV} along with a methodology to assess object storage performance for NWP applications, which is also adopted here. 
Following that methodology, the IOR benchmark has been configured so that each I/O process issues a single, large, I/O operation comprising a sequence of data parts (we refer to this mode of operation as \textit{segments} mode). This enables assessment of the maximum achievable performance if the developed application were optimised to gather and transfer all relevant data in a single I/O operation, and provides insight on to what extent the storage server is able to exploit available network bandwidth and/or storage capability when not having to deal with new operations.

In order to execute the object-store-oriented Field I/O benchmark against POSIX Lustre, a helper library has been developed which implements the DAOS API using POSIX file system concepts. DAOS Pools have been implemented as directories hosted on the distributed file system. Containers have been implemented as directories under the corresponding pool directories. Key-Value objects have been implemented as directories under the corresponding container directory. A key is implemented as a file in the Key-Value directory, named with the key name, and a value is stored as content in that key file. Array objects are implemented as files under the corresponding container directory, named with the object ID of the Array and containing the array data.

%diagram?

Following this design, every write of a weather field performed by the Field I/O benchmark using the helper library will usually involve: a) write of an Array file in a Container directory exclusive for every client process, b) check existence of a Key-Value directory in a Container directory shared with all processes in the client node, c) creation and write of a Key file in a Key-Value directory shared with all processes in the client node.

A field read will usually involve: a) check existence of a Key-Value directory in a Container directory shared with all client processes, b) open and read of a Key file in a Key-Value directory shared by all client processes, c) check existence of a Key-Value directory in a Container directory shared with all processes in the client node, d) open and read of a Key file in a Key-Value directory shared with all processes in the client node, e) open and read an Array file in a Container directory exclusive for the process.

Whereas the bandwidth metric used to quantify performance in IOR runs has been the one provided by IOR itself, referred to as \textit{synchronous bandwidth} here, a custom bandwidth metric for non-synchronised applications is used in Field I/O runs, the \textit{global timing bandwidth}. Both metrics are described in\cite{DAOS-IEEETPDS-ARXIV}.

\subsection{NEXTGenIO}

The benchmarks we present have been conducted in NEXTGenIO\cite{BAPM-isc2019}, a research HPC system composed of 34 dual-socket nodes with Intel Xeon Cascade Lake processors. Each socket has six 256 GiB first-generation Intel's Optane Data Centre Persistent Memory Modules (DCPMMs) configured in AppDirect interleaved mode, although there are no NVMe devices. Each processor is connected with its own integrated network adapter to a low-latency OmniPath fabric. Each of these adapters has a maximum bandwidth of 12.5 GiB/s. %The fabric is configured in dual-rail mode, that is, two separate OmniPath switches interconnect first-socket adapters and second-socket adapters separately, respectively.

For the Lustre benchmarking, Lustre has been deployed on 8 storage nodes (providing 16 OSTs, one per socket), plus one node devoted to the metadata service. Both the OSTs and the MDTs used in the file system mount an ext4 file system on the SCM attached to their respective sockets, providing 1.5 TB of high-performance storage per OST and MDT.

To run the benchmarks against configurable amounts of Lustre storage nodes, Lustre pools have been set up with 1, 2, 4 and 8 nodes. Before running a test, a folder is created in the file system where all test files will be generated, and the \verb!setstripe! command is used to bind that folder to the pool with the desired number of storage nodes.

DAOS deployments have been conducted separately in an ad-hoc basis, removing and re-deploying with the desired amount of storage nodes as needed. Each node used for DAOS storage deploys a single DAOS engine per socket, using the full ext4 file system on the Optane SCM for that socket, and has access to all the cores available on the associated processor. For instance, to compare to a two-node Lustre configuration (which uses two OST nodes and one MDT node, giving 4 OSTs overall), a two-node DAOS deployment is created and provisioned for the benchmark.

Up to 16 nodes were employed to execute the benchmark client processes using both sockets and network interfaces. SCM in the client nodes was not used and did not have any effect on I/O performance.

\section{Results}

Performance results obtained from running the described benchmarks against Lustre and DAOS deployments in NEXTGenIO are discussed next, starting with an assessment of potential achievable performance with IOR, following with real achieved performance with Field I/O without and with contention between writer and reader applications, and concluding with an assessment of the impact of file or object size in such applications.

\subsection{IOR results, potential performance}

In this test set, the IOR benchmark has been run against different amounts of Lustre and DAOS storage server nodes with the intent to analyse maximum write and read performance a client application can potentially achieve, and to analyse the behaviour of the storage performance as more server nodes are added. IOR has been run in segments mode as described above, following access pattern A.

For each server node count the benchmark has been run using as many client nodes, twice as many and, where possible, four times as many, with the aim to effectively make use of the available bandwidth theoretically provided by the server network interfaces. For each of these combinations, the benchmark has been run with 36, 48, 72 and 96 IOR processes per client node as these were found to result in the best performance in preliminary tests, for both Lustre and DAOS. Each run has been repeated 5 times to account for variability.

The segment count has been set to 100 as it was found to be the minimum segment count that results in reduced bandwidth variability. The segment size has been set to 1 MiB, to match the object or field size in the NWP use case. This segment configuration results in files or objects of 100 MiB in size being written and read during the benchmark runs.

%DAOS and Lustre striping have been disabled to avoid complexity in network behaviour. <-- not really! Lustre striping was set to 1MiB default stripe size and, together with -c -1, the 100MiB files were striped across all storage nodes

The results are shown in Fig. \ref{fig:ior}. Each dot represents the mean synchronous bandwidth obtained with the best performing number of client nodes and IOR process count for the corresponding server node count. Hollow dots are used to indicate cases where it was only possible to use up to twice as many client nodes as server nodes, not  four times as many as in the rest of the benchmarks.

\begin{figure}[htbp]
    \begin{subfigure}[b]{124pt}
        \includegraphics[width=124pt]{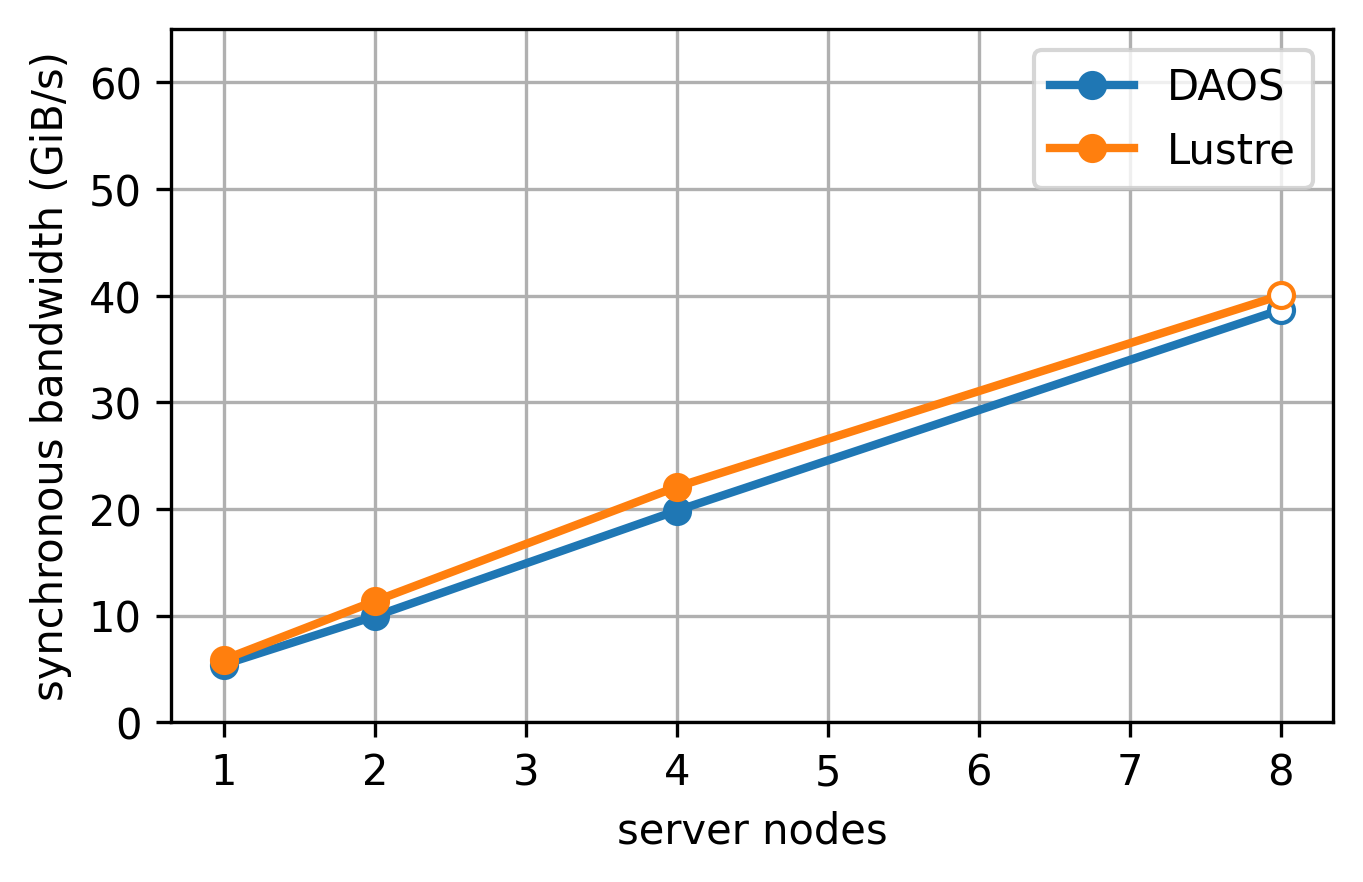}
        \caption{Write}
    \end{subfigure}
    \begin{subfigure}[b]{124pt}
        \includegraphics[width=124pt]{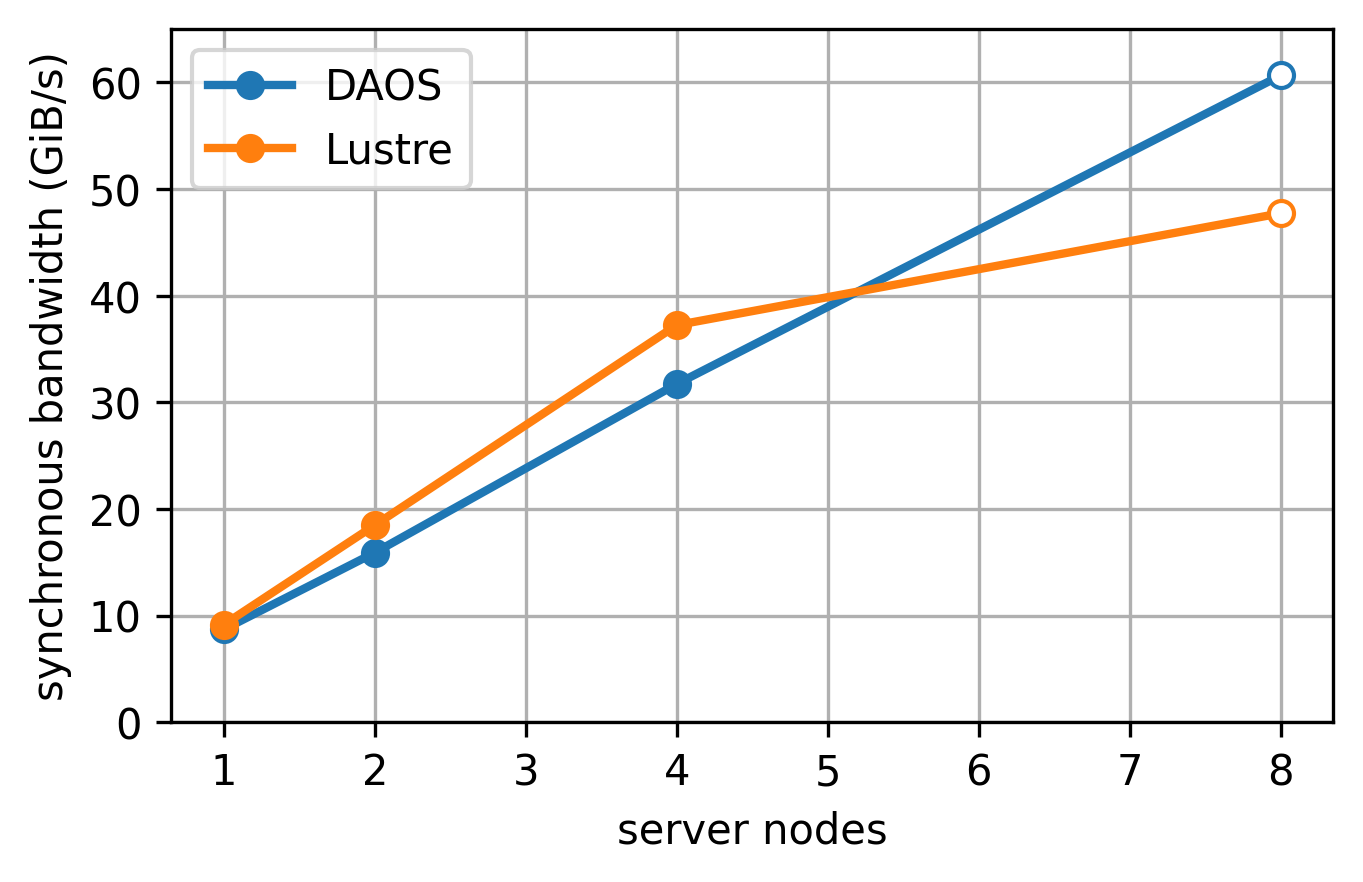}
        \caption{Read}
    \end{subfigure}
    \caption{Mean synchronous write and read bandwidth results for Lustre and DAOS for access pattern A with IOR in segments mode.}
    \label{fig:ior}
\end{figure}
%\begin{figure*}[htbp]
%    \centering
%    \begin{subfigure}[b]{250pt}
%        \centering
%        \includegraphics[width=250pt]{results-ior_segments_scaling_lustre_vs_daos_write_mean.png}
%        \caption{Write}
%    \end{subfigure}
%    %\vskip\baselineskip
%    \begin{subfigure}[b]{250pt}
%        \centering
%        \includegraphics[width=250pt]{results-ior_segments_scaling_lustre_vs_daos_read_mean.png}
%        \caption{Read}
%    \end{subfigure}
%    \caption{Mean synchronous write and read bandwidth results for Lustre and DAOS for access pattern A (unique writes then unique reads) with IOR in segments mode.}
%    \label{fig:ior}
%\end{figure*}

Lustre and DAOS perform similarly, resulting in comparable write and read benchmark bandwidths, with a similar scaling pattern. These results indicate that both storage servers and the benchmarks have been properly configured to exploit available storage and network resources, which has been further verified with monitoring of resource usage during preliminary test runs.

Slightly higher bandwidths are achieved overall with Lustre in this scenario, except in the read phase in the configuration with 8 server nodes, which was tested with only up to twice as many client nodes running the benchmark. The performance limitation in that configuration is explained by the fact that, in our test platform and with our configuration, four times as many client nodes as Lustre server nodes are required to exploit all available server interface bandwidth whereas, with DAOS, only twice as many client nodes are required. Results showing optimal client to server node ratio with Lustre have been omitted in favour of space. For DAOS, this ratio is addressed in\cite{DAOS-IEEETPDS-ARXIV}.

Excluding this special case, write and read IOR bandwidths obtained with both Lustre and DAOS scale linearly. For Lustre, the benchmark bandwidth increases at a rate of approximately 6 GiB/s for write and 9 GiB/s for read per additional server node. For DAOS, the increase is of approximately 5 GiB/s for write and 7.5 GiB/s for read.

\subsection{Field I/O results}

\subsubsection{Application performance with DAOS vs. Lustre}

The Field I/O benchmark, in contrast to the IOR benchmark in segments mode, entails object-store-like I/O operations. The data elements are small, and writing or reading a single data object usually involves multiple I/O operations.

In this test set, Field I/O has been run with access patterns A and B following a similar strategy to the IOR tests above, using both Lustre and DAOS with the similar amounts of server and client nodes. In the runs here, a maximum of twice as many client nodes as server nodes have been employed for all configurations.

The benchmark has been run in its default \textit{full} mode, and also in a mode which avoids use of DAOS containers called \textit{no-containers}\cite{DAOS-IEEETPDS-ARXIV}. When run against Lustre, using the helper library to adapt Field I/O to POSIX file systems, every field write in the benchmark in no-containers mode will involve less metadata operations but the Array files will be written to a single directory shared by all client processes. Field reads will equally read Array files from that shared directory.

The object size has been set to 1 MiB, and the number of I/O iterations per process to 2000, to reduce the impact of potential parallel process start-up delays on bandwidth measurements. The chosen amounts of processes per client node, which have been found to result in best performance, have been 24 and 36 (slightly lower than the 36 and 48 found for DAOS). Each run has been repeated 5 times.

Results for pattern A and B are shown in Fig. \ref{fig:fieldio_pattern_a} and Fig. \ref{fig:fieldio_pattern_b}, respectively. Note that, in both figures, Lustre results on the left side and DAOS results on the right side use a different y-axis scale.

\begin{figure*}[htbp]
    \centering
    \begin{subfigure}[b]{125pt}
        \includegraphics[width=125pt]{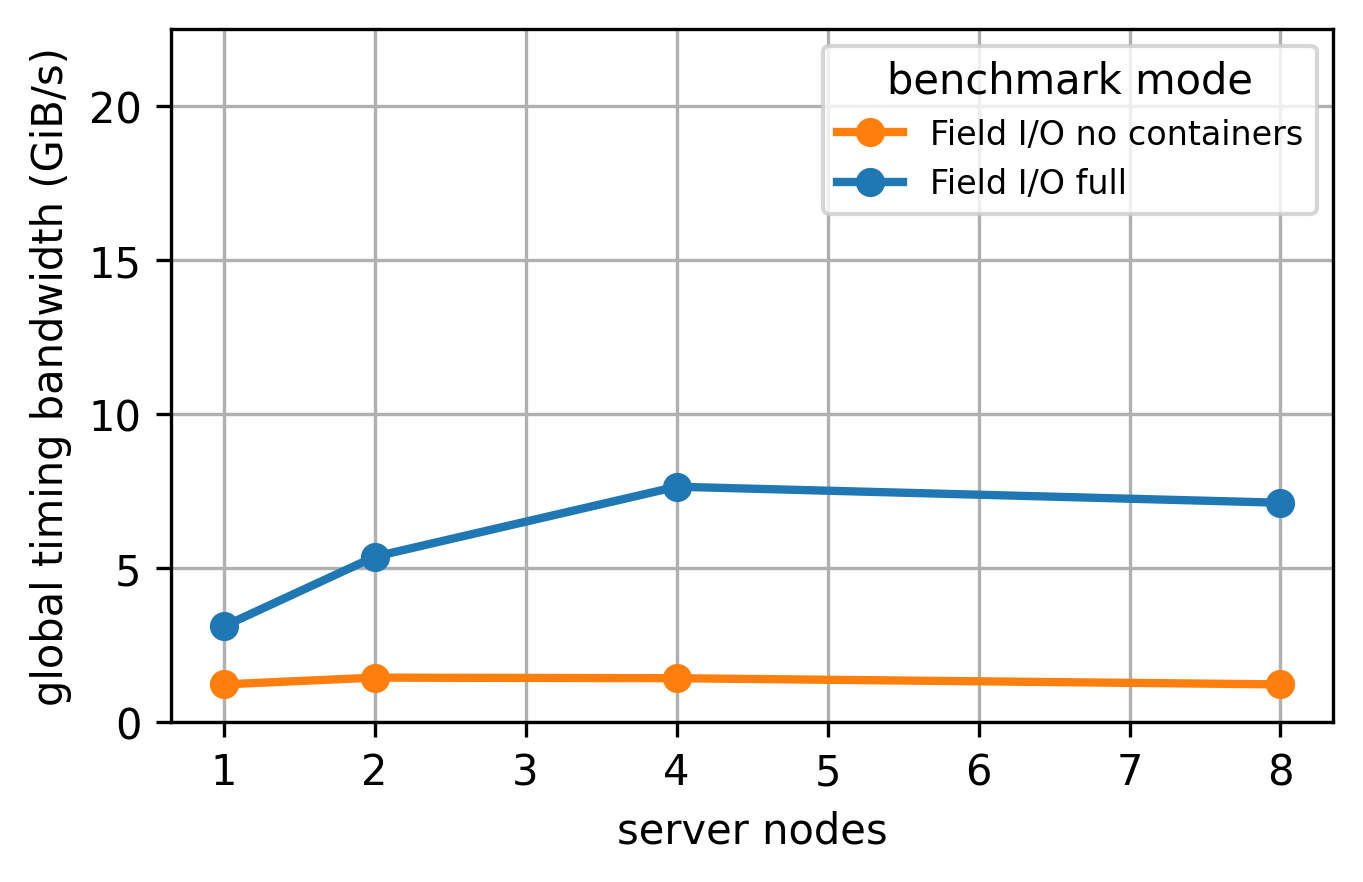}
        \caption{Lustre, write}
    \end{subfigure}
    \begin{subfigure}[b]{125pt}
        \includegraphics[width=125pt]{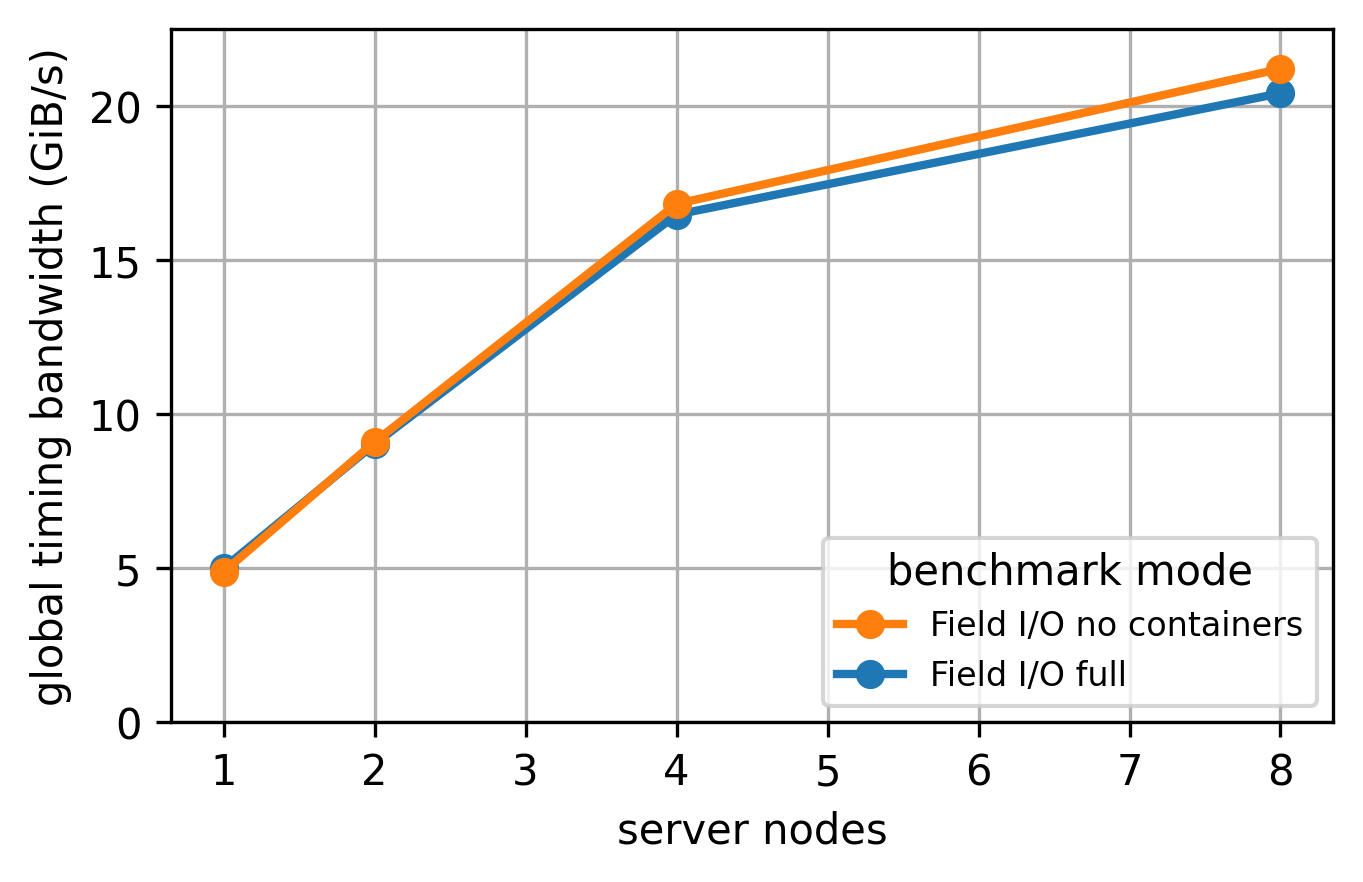}
        \caption{Lustre, read}
    \end{subfigure}
    \begin{subfigure}[b]{125pt}
        \includegraphics[width=125pt]{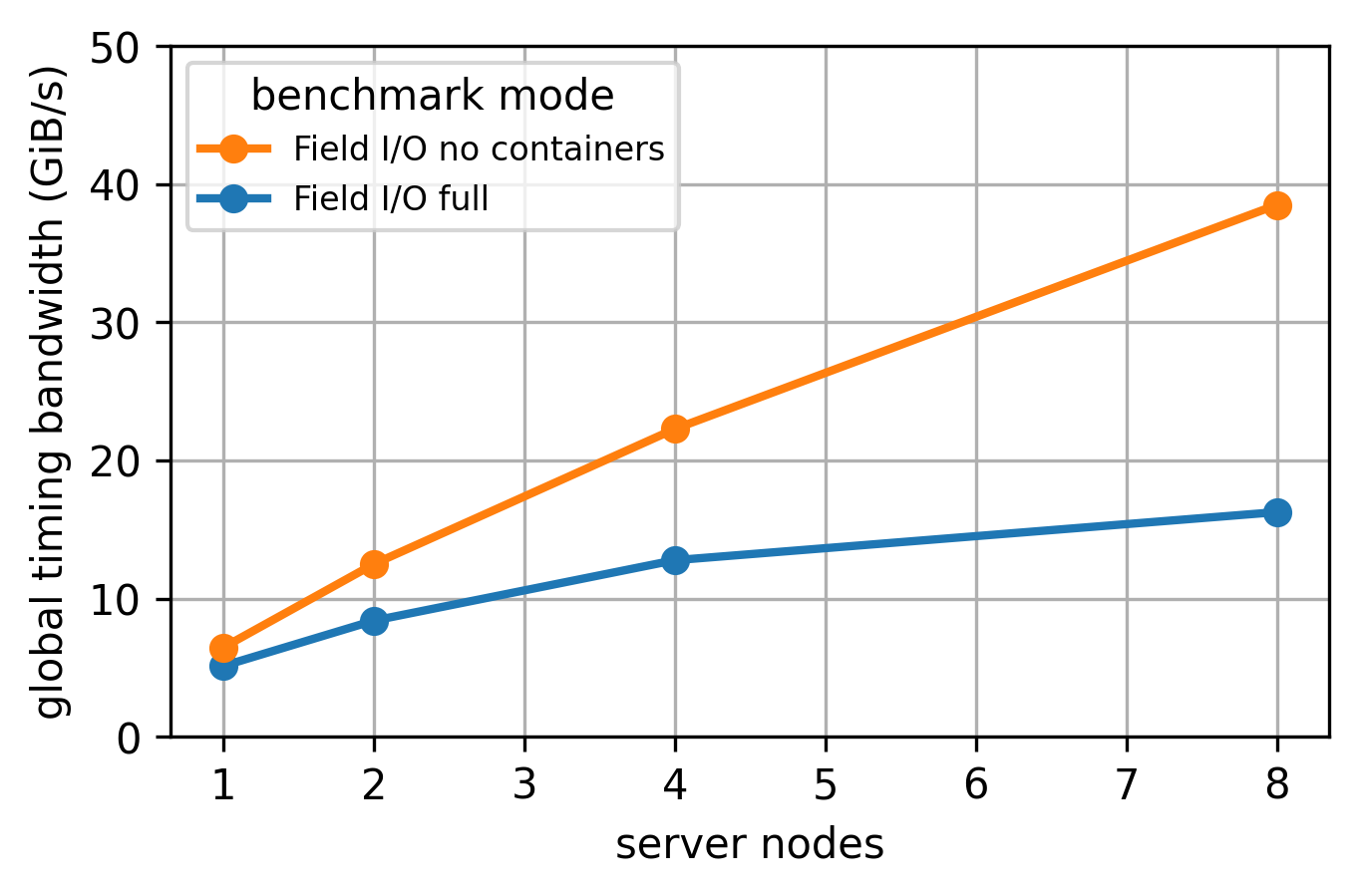}
        \caption{DAOS, write}
    \end{subfigure}
    %\vskip\baselineskip
    \begin{subfigure}[b]{125pt}
        \includegraphics[width=125pt]{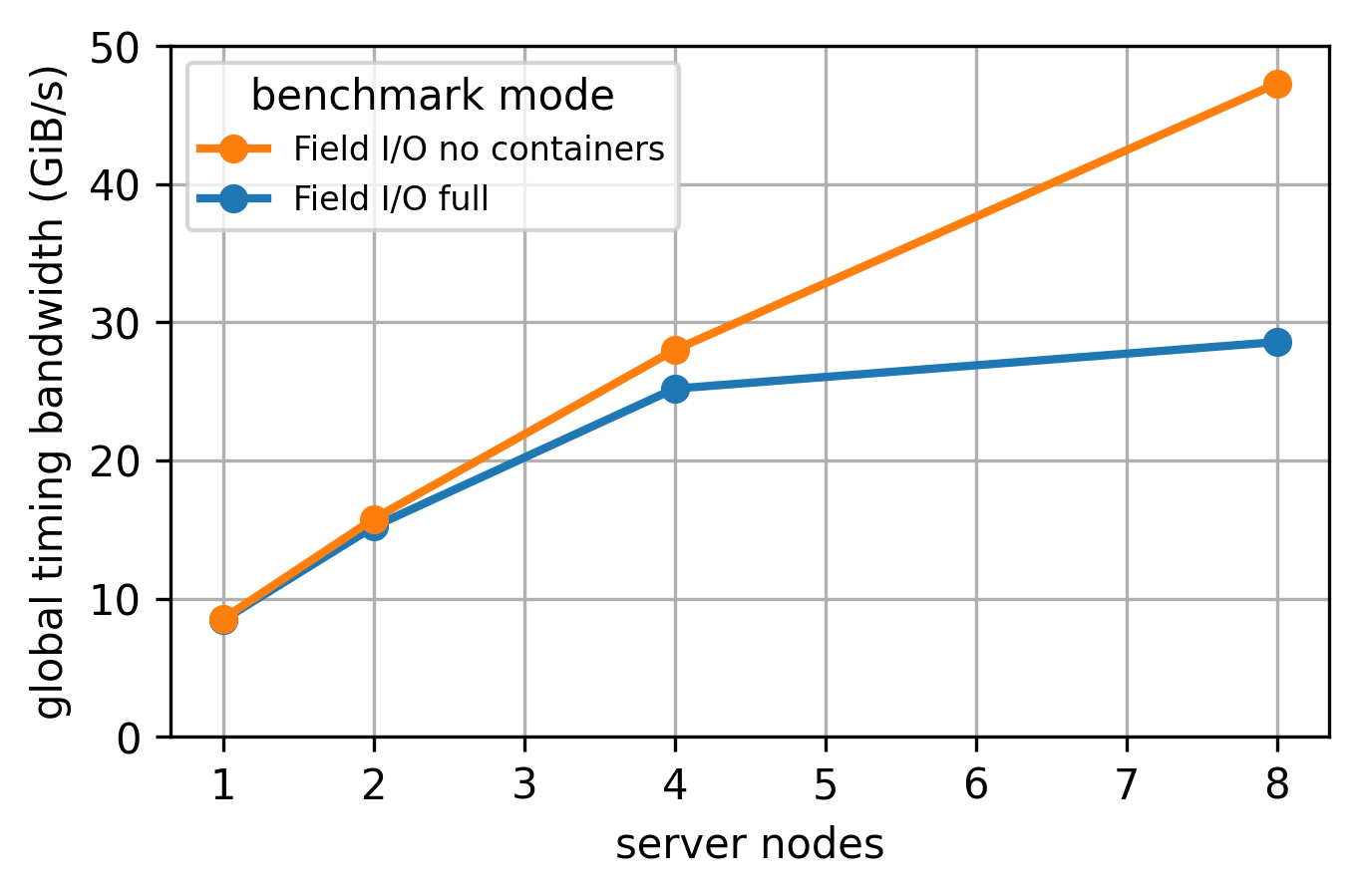}
        \caption{DAOS, read}
    \end{subfigure}
    \caption{Access pattern A, global timing write and read bandwidth results with the Field I/O benchmark.}% The bandwidth scales very well, particularly if multiple containers are avoided.}
    \label{fig:fieldio_pattern_a}
\end{figure*}

\begin{figure*}[htbp]
    \centering
    \begin{subfigure}[b]{125pt}
        \includegraphics[width=125pt]{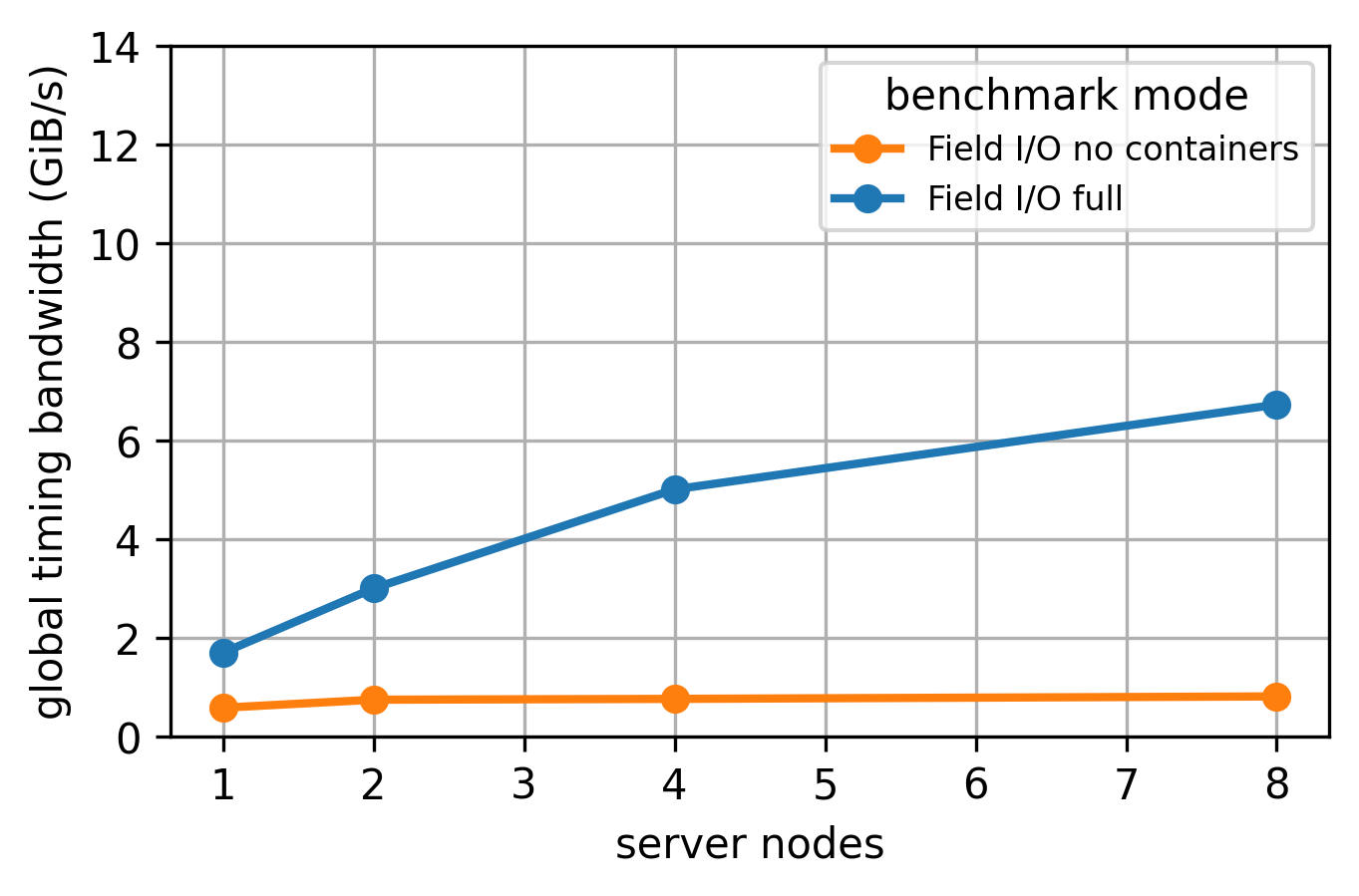}
        \caption{Lustre, write}
    \end{subfigure}
    \begin{subfigure}[b]{125pt}
        \includegraphics[width=125pt]{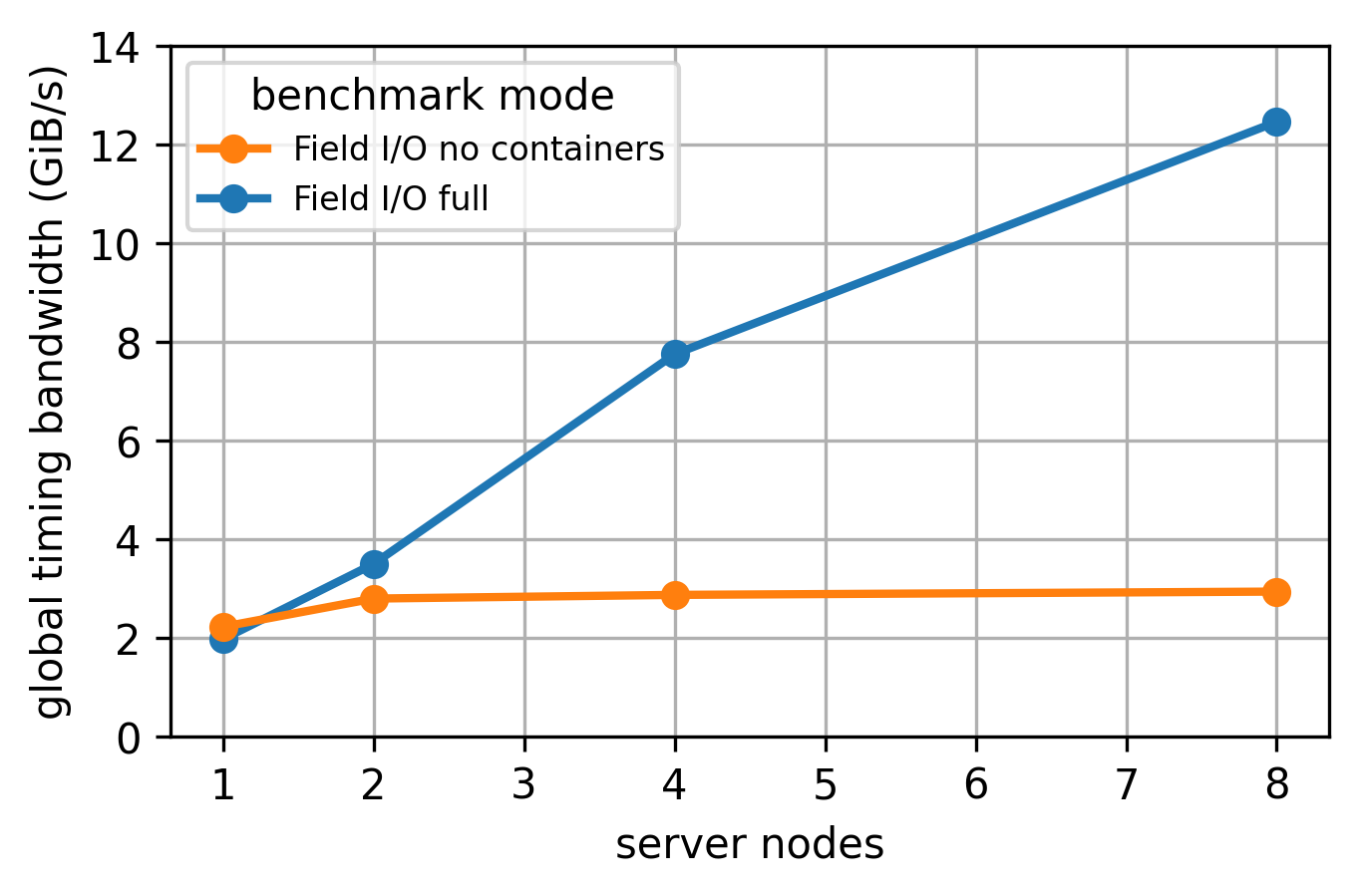}
        \caption{Lustre, read}
    \end{subfigure}
    \begin{subfigure}[b]{125pt}
        \includegraphics[width=125pt]{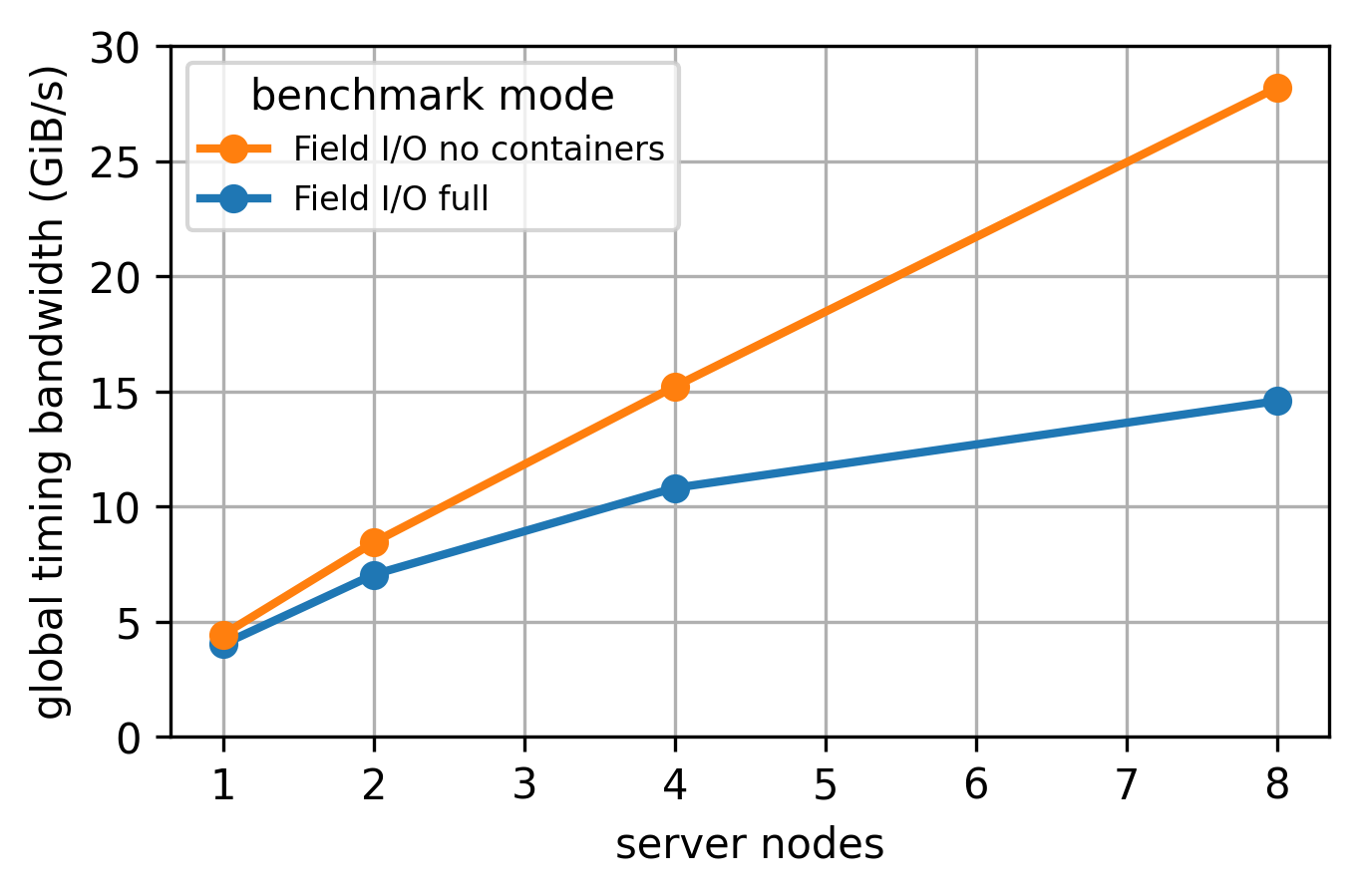}
        \caption{DAOS, write}
    \end{subfigure}
    %\vskip\baselineskip
    \begin{subfigure}[b]{125pt}
        \includegraphics[width=125pt]{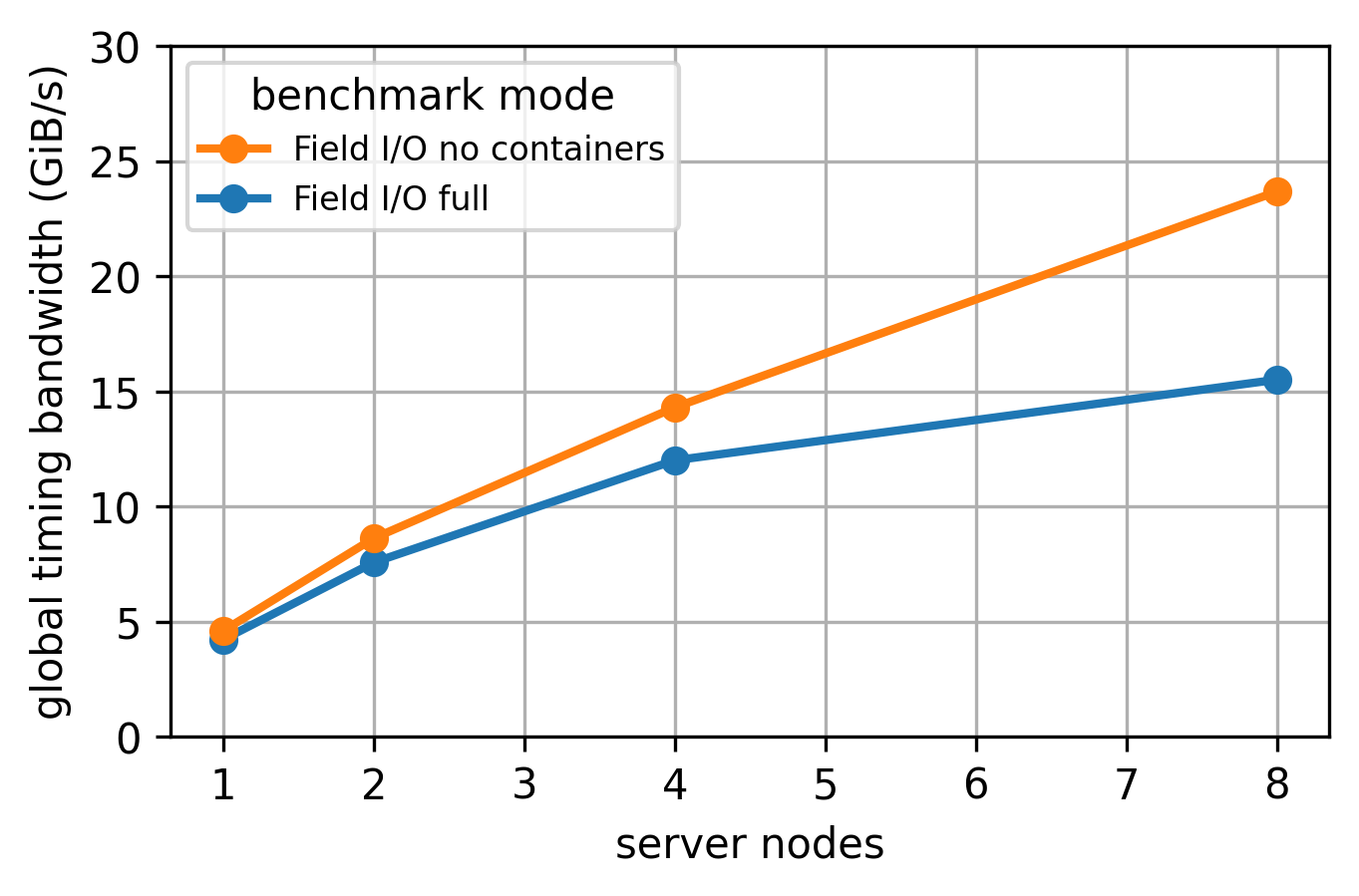}
        \caption{DAOS, read}
    \end{subfigure}
    \caption{Access pattern B, global timing write and read bandwidth results with the Field I/O benchmark.}% The bandwidth scales very well, particularly if multiple containers are avoided.}
    \label{fig:fieldio_pattern_b}
\end{figure*}

Looking at DAOS results for pattern A, in Fig. \ref{fig:fieldio_pattern_a} (c) and (d), we can see that Field I/O in no-containers mode performs better than the mode with containers. This is possibly due to inefficiencies in the use of DAOS containers, as discussed in \cite{DAOS-IEEETPDS-ARXIV}. The mode without containers scales linearly, and the application bandwidth increases at an approximate rate of 4.5 GiB/s write and 5.5 GiB/s read per additional server node.

Lustre results in Fig. \ref{fig:fieldio_pattern_a} (a) show that Field I/O in no-containers mode, using the helper library to adapt to POSIX, performs poorly for write. This is likely due to performing all the Array file writes and reads in a single directory, as explained above, which suffers from Lustre contention or locking on that directory.

The full mode, which has the Array files distributed in several Container directories, performs better and reaches up to 7.5 GiB/s for write, but hits a limit at 4 server nodes. In benchmark runs with more server nodes, more client nodes are also employed, and every client node is set to execute an additional fixed amount of I/O operations. The limitation observed here is due to reaching to IOPs limits on the Lustre metadata server, which we have benchmarked at around 100 KIOPs using IOR. Measured IOPs rates for the Field I/O benchmark runs approach this limit where the scaling limit is reached. In IOR benchmarks Lustre can scale up to much larger bandwidths because there are less IOPs are involved.

Fig. \ref{fig:fieldio_pattern_a} (b) shows that both Field I/O modes perform similarly for read, and they similarly hit a limit beyond 4 server nodes, again due to reaching maximum IOPs on the Lustre metadata server. We postulate that both modes provide similar performance due to the lack of locking/coherency required for read operations in Lustre.

For results in Fig. \ref{fig:fieldio_pattern_b} for access pattern B, where writer applications are run concurrently with reader applications, the write and read bandwidths need to be combined for an approximate comparison with results with pattern A. With DAOS, Field I/O in no-containers mode performs remarkably well, with linear scaling and an aggregated bandwidth of approximately 50 GiB/s with 8 server nodes, higher than the separate bandwidths in pattern A.

With Lustre, the write and read performance behaviour of the full mode with access pattern B is similar to that observed in pattern A, with slightly better scaling beyond 4 nodes. The no-containers mode performs poorly not only for write but also for read, likely due to locking/coherency issues caused by concurrent writes.

In terms of achieved aggregate application bandwidth, Lustre with Field I/O pattern B in full mode reaches up to approximately 20 GiB/s with 8 server nodes, close to the read bandwidth obtained in pattern A, but far from the higher aggregate bandwidth obtained with DAOS with pattern B in any of the modes.

From the performance results for access pattern A, it can be observed that the best application bandwidths obtained with DAOS are in the same order of magnitude as those obtained with IOR, whereas with Lustre they are approximately a fifth of those obtained with IOR.

We have identified four factors that could contribute to this performance difference:

\begin{itemize}[leftmargin=*]
    \item Lustre is designed for large file I/O, and has a bottleneck in the metadata server when exposed to object-store-like applications.
    \item the Field I/O application, when run against Lustre, uses the helper library to adapt to POSIX, which implements a custom indexing mechanism with Key-Value directories and files. An implementation that made use of Lustre's own directory and file names for indexing would involve less file metadata operations and may perform better.
    \item Lustre needs four times as many client nodes as server nodes to saturate network bandwidth, but we have run the benchmarks with only twice as many.
    \item DAOS has been optimised for new memory technologies such as SCM and can bypass the block storage interface for some operations.
\end{itemize}

\subsubsection{Impact of object and file size}

Fig. \ref{fig:object_size} shows bandwidth results from runs of access pattern A with Field I/O in full mode with Lustre, and no-containers mode with DAOS. This time the benchmark has been run with varying object sizes of 1, 5, 10 and 20 MiB, to assess the impact of future NWP model resolution increases on I/O performance.

All tests have been run with a fixed configuration of 2 server nodes and 4 client nodes, and 100 I/O operations per client process. Benchmarks are repeated 5 times, with the same client process counts as in previous Field I/O runs. The results for the 5 repetitions have been averaged, with the average bandwidth for the highest performing number of client processes per client node shown for each I/O size tested.

% As previously discussed, there are numerous ways that DAOS can be configured when creating the system or containers within the system. One of the key user mechanisms for configuration is setting the object class associated with a given object. We investigated varying object classes, which can be considered similar to striping of data in a system such as Lustre, or distributing objects to different DAOS engines, depending on the object sizes. For this benchmarking we also varied the object size, to evaluate the performance impact of I/O size on DAOS.
%In this test set we investigated the effect of object granularity and different striping configurations for the Field I/O objects.

%Fig. \ref{fig:object_class_bandwidth} shows bandwidth results from runs of access pattern A with Field I/O in full mode, with object sizes 1, 5, 10 and 20 MiB, and object configurations ranging from no striping (\verb!OC_S1!) to striping all objects across all targets (\verb!OC_SX!). Field I/O has been configured to run with high contention in the indexing Key-Values (i.e. Field I/O full mode from Fig. \ref{fig:fieldio}).

\begin{figure}[htbp]
    \begin{subfigure}[b]{124pt}
        \includegraphics[width=124pt]{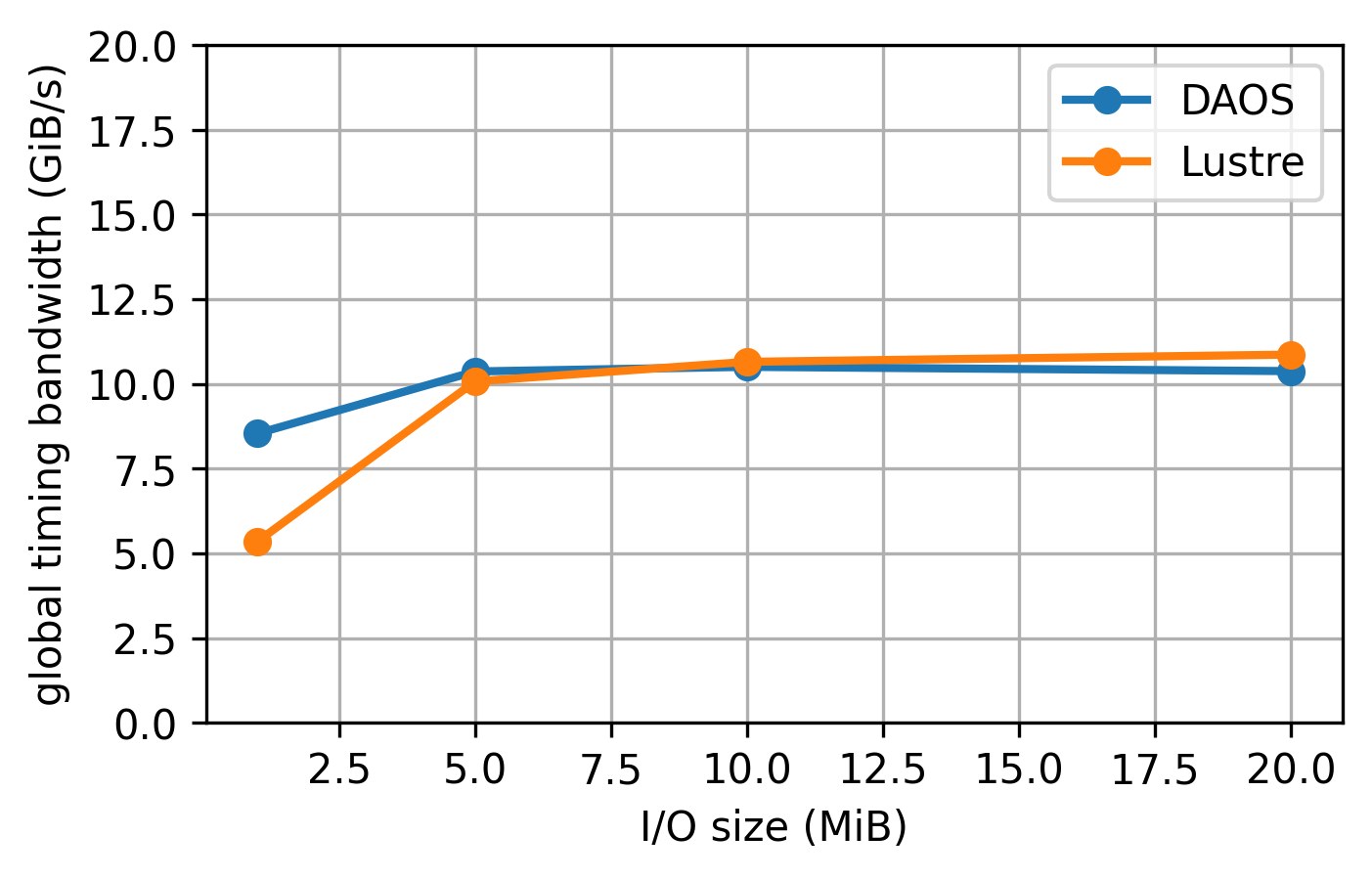}
        \caption{Write}
    \end{subfigure}
    \begin{subfigure}[b]{124pt}
        \includegraphics[width=124pt]{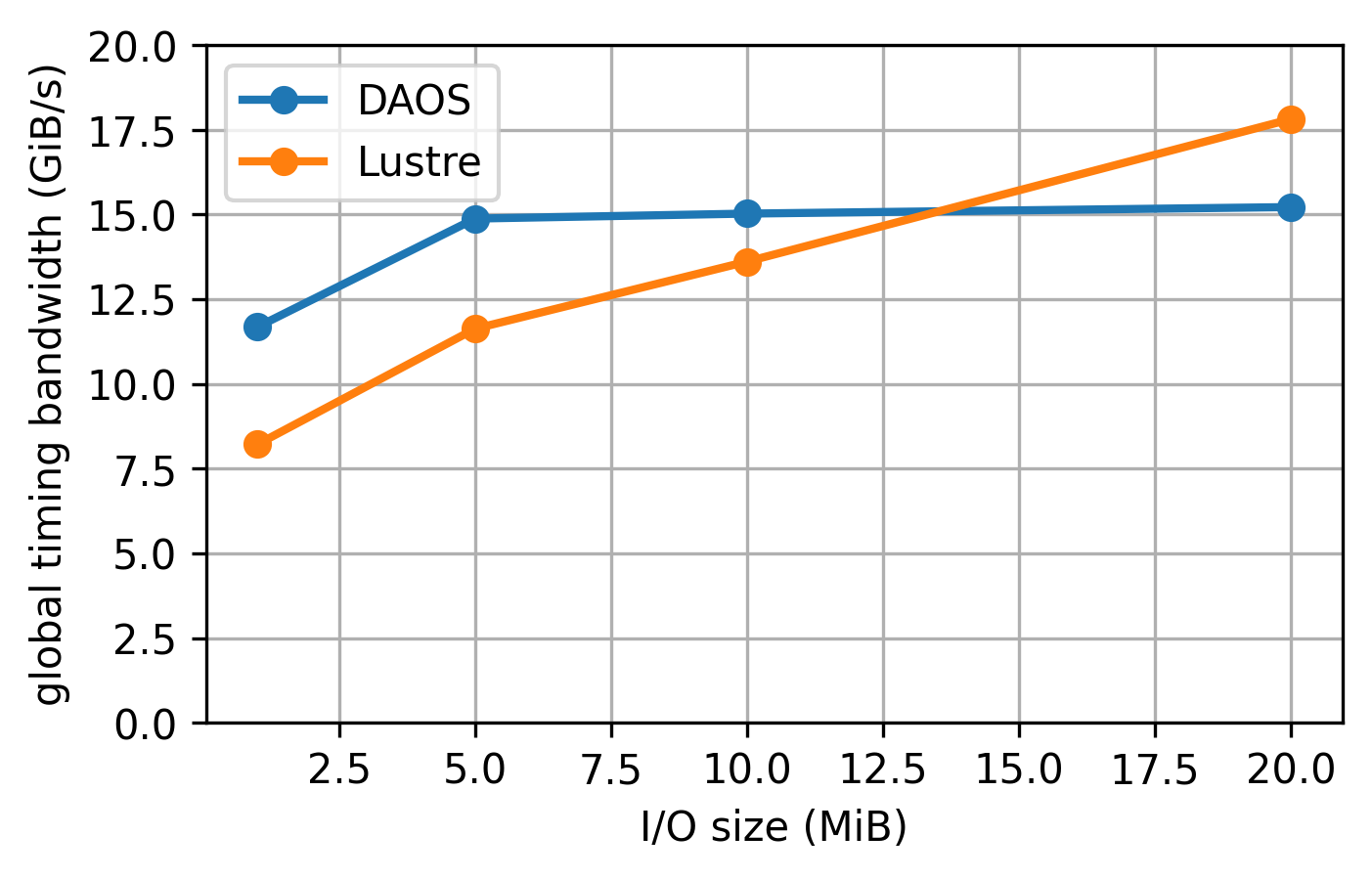}
        \caption{Read}
    \end{subfigure}
    \caption{Global timing write and read bandwidth results for access pattern A, with the Field I/O benchmark, using 2 server nodes and 4 client nodes.}
    \label{fig:object_size}
\end{figure}
%\begin{figure*}[htbp]
%    \centering
%    \begin{subfigure}[b]{250pt}
%        \centering
%        \includegraphics[width=250pt]{results-osize-lustre_vs_daos_write.png}
%        \caption{Write}
%    \end{subfigure}
%    %\vskip\baselineskip
%    \begin{subfigure}[b]{250pt}
%        \centering
%        \includegraphics[width=250pt]{results-osize-lustre_vs_daos_read.png}
%        \caption{Read}
%    \end{subfigure}
%    \caption{Global timing write and read bandwidth results for access pattern A (unique writes then unique reads), with the Field I/O benchmark, using 2 server nodes and 4 client nodes.}
%    \label{fig:object_size}
%\end{figure*}

It can be observed that increasing object or file size from 1 MiB to 5 MiB has a substantial benefit for both write and read, with both Lustre and DAOS. Beyond 5 MiB, the bandwidth stabilizes except in Lustre reading, where it continues to increase.

In this benchmark configuration, the bandwidths obtained with DAOS are higher than with Lustre when using an object or file size of 1 MiB. The two storage systems provide similar write bandwidths for larger object or file sizes. For read, DAOS performs better than Lustre with objects up to 10 MiB in size, and Lustre performs better than DAOS with larger files.

This test set could be repeated using access pattern B to investigate whether application bandwidth behaves differently as object or file size increase under write and read contention. Likewise, different server and client node configurations could be investigated as well.

\section{Related Work}
There has been research into object stores for high-performance I/O, including CEPH \cite{ceph-pdp19} and CORTX Motr \cite{mero-cf18}. However, these have so far seen less adoption for very intensive data creation or processing workloads in large-scale systems. DAOS is one of the first production-ready object storage technologies targeting HPC, with remarkable results in recent IO-500 benchmarks\cite{io500-isc2021}.

Science communities have investigated object stores for I/O operations\cite{fdb-pasc17}, including using non-volatile memory\cite{fdb-pasc19}. These investigations have demonstrated the performance and functionality benefits, however they highlight direct porting relies on custom management of objects on storage mediums. Domain-agnostic object stores like DAOS simplify the use of NVM in production environments.

The work presented in this paper builds on previous research in the areas of exploiting object storage and NVM technologies for HPC I/O, extending understanding and knowledge on the impact of the object store approach versus the benefits of using high-performance I/O hardware. It also extends community understanding of how DAOS and Lustre perform under workload patterns common in many applications.

\section{Conclusions}

Through our benchmarking and evaluation work we have demonstrated that Lustre and DAOS, using the same underlying storage hardware, can provide similar performance for large-scale bulk data operations, such as those the IOR benchmark mimics and many applications have been tuned to utilise. However, when moving to object-store-like I/O operations and workflows, DAOS shows significant performance benefits. We conclude that DAOS is a strong target for a future storage platform as it demonstrates comparative performance with traditional files systems whilst enabling object-store functionality at high-performance.

The access patterns produced with the Field I/O benchmark are not particularly specific to the NWP use case or bound to any particular file format, and therefore the findings are applicable to other use cases repeatedly writing and/or reading any objects or files with sizes in the order of 1 to 10 MiB from several processes and nodes.
%DAOS has shown better performance than Lustre for object-store-like I/O workflows, with small-sized writes and reads and more metadata operations. For standard file system-like operations, DAOS has achieved similar performance to Lustre employing less client nodes. Therefore moving to DAOS for standard applications should not degrade performance, and it should enable high-performance data object operations.

The results suggest that the good performance results obtained with DAOS in the object-store-like NWP use case are not purely due to the use of high-performance storage hardware such as SCM, but also due to the design characteristics of an object store and their implementation in DAOS.

Nevertheless, performance and scalability of Lustre in this scenario should be further explored and validated by benchmarking with larger amounts of server and client nodes, and possibly implementing additional optimisations in the benchmarks to better exploit file system capabilities.

\section*{Acknowledgment}

The work presented in this paper has been produced in the context of the European Union’s Destination Earth Initiative and relates to tasks entrusted by the European Union to the European Centre for Medium-Range Weather Forecasts implementing part of this Initiative with funding by the European Union (cost center DE3100, code 3320). Views and opinions expressed are those of the author(s) only and do not necessarily reflect those of the European Union or the European Commission. Neither the European Union nor the European Commission can be held responsible for them. This work was also supported by EPSRC, grant EP/T028351/1.

The NEXTGenIO system was funded by the European Union's Horizon 2020 Research and Innovation program under Grant Agreement no. 671951, and supported by EPCC, The University of Edinburgh.


\begin{thebibliography}{00}
\bibitem{nextplatform_lockwood} G. Lockwood, "What's so bad about POSIX I/O?". The Next Platform 2017. https://www.nextplatform.com/2017/09/11/whats-bad-posix-io/
\bibitem{fdb-pasc19} S. Smart, T. Quintino, and B. Raoult. "A High-Performance Distributed Object-Store for Exascale Numerical Weather Prediction and Climate". In Proceedings of the Platform for Advanced Scientific Computing Conference (PASC '19). Association for Computing Machinery, New York, NY, USA, Article 16, 1–11. DOI:https://doi.org/10.1145/3324989.3325726
\bibitem{DAOS-IEEETPDS-ARXIV} N. Manubens, T. Quintino, S. D. Smart, E. Danovaro, and A. Jackson, “DAOS as HPC Storage, a view from Numerical Weather Prediction”, arXiv e-prints, https://arxiv.org/abs/2208.06752, 2022.
\bibitem{daos-scfa2020} Z. Liang, J. Lombardi, M. Chaarawi, M. Hennecke, "DAOS: A Scale-Out High Performance Storage Stack for Storage Class Memory" In: Panda, D. (eds) Supercomputing Frontiers. SCFA 2020. Lecture Notes in Computer Science(), vol 12082. Springer, Cham. DOI:https://doi.org/10.1007/978-3-030-48842-0\_3
\bibitem{lustre-arxiv2019} P. Braam "The Lustre Storage Architecture" 2019 arXiv. DOI:https://doi.org/10.48550/arXiv.1903.01955
\bibitem{ior_repo} "HPC IO Benchmark Repository", (2022), GitHub repository, https://github.com/hpc/ior
\bibitem{BAPM-isc2019} A. Jackson, M. Weiland, M.Parsons, and B. Homölle. "An Architecture for High Performance Computing and Data Systems Using Byte-Addressable Persistent Memory". In High Performance Computing: ISC High Performance 2019 International Workshops, Frankfurt, Germany, June 16-20, 2019. Springer-Verlag, Berlin, Heidelberg, 258–274. DOI:https://doi.org/10.1007/978-3-030-34356-9\_21
\bibitem{BAPM-sc2019} M. Weiland, H. Brunst, T. Quintino, et al. "An early evaluation of Intel's optane DC persistent memory module and its impact on high-performance scientific applications". In Proceedings of the International Conference for High Performance Computing, Networking, Storage and Analysis . Association for Computing Machinery, New York, NY, USA, Article 76, 1–19. DOI:https://doi.org/10.1145/3295500.3356159
\bibitem{ceph-pdp19} K. Jeong, C. Duffy, J. -S. Kim and J. Lee, "Optimizing the Ceph Distributed File System for High Performance Computing," 2019 27th Euromicro International Conference on Parallel, Distributed and Network-Based Processing (PDP), 2019, pp. 446-451, DOI: 10.1109/EMPDP.2019.8671563.
\bibitem{mero-cf18} S. Narasimhamurthy, N. Danilov, S. Wu, G. Umanesan, S. Chien, S. Rivas-Gomez, I. Peng, E. Laure, S. Witt, D. Pleiter, and S. Markidis. "The SAGE project: a storage centric approach for exascale computing: invited paper". In Proceedings of the 15th ACM International Conference on Computing Frontiers (CF '18). Association for Computing Machinery, New York, NY, USA, 287–292. DOI:https://doi.org/10.1145/3203217.3205341
\bibitem{io500-isc2021}  A. Dilger, D. Hildebrand, J. Kunkel, J. Lofstead, and G. Markomanolis, "IO500 10 node list Interational Supercomputing 2021, June 2021, [online] Available:https://doi.org/10.5281/zenodo.5171694
\bibitem{fdb-pasc17} S. Smart, T. Quintino, and B. Raoult "A Scalable Object Store for Meteorological and Climate Data". In Proceedings of the Platform for Advanced Scientific Computing Conference (PASC '17). Association for Computing Machinery, New York, NY, USA, Article 13, 1–8. DOI:https://doi.org/10.1145/3093172.3093238
\bibitem{gekko-jcst2020} MA. Vef, N. Moti, T. Süß et al. "GekkoFS — A Temporary Burst Buffer File System for HPC Applications". J. Comput. Sci. Technol. 35, 72–91 (2020). https://doi.org/10.1007/s11390-020-9797-6
\bibitem{chfs-hpcasia2022} O. Tatebe, K. Obata, K. Hiraga, and H. Ohtsuji. 2022. "CHFS: Parallel Consistent Hashing File System for Node-local Persistent Memory". In International Conference on High Performance Computing in Asia-Pacific Region (HPCAsia2022). Association for Computing Machinery, New York, NY, USA, 115–124. DOI:https://doi.org/10.1145/3492805.3492807
\bibitem{norns-ieeecluster2019} A. Miranda, A. Jackson, T. Tocci, I. Panourgias and R. Nou, "NORNS: Extending Slurm to Support Data-Driven Workflows through Asynchronous Data Staging" 2019 IEEE International Conference on Cluster Computing (CLUSTER), 2019, pp. 1-12, doi: 10.1109/CLUSTER.2019.8891014.
\end{thebibliography}
\end{document}